\documentclass[12pt,aps,showpacs,eqsecnum,nofootinbib,floatfix]{revtex4}
\usepackage{bbding}
\usepackage{pifont}
\usepackage{subfigure}
\usepackage{epsfig}
\usepackage{array}
\usepackage{amsmath}
\usepackage{amsfonts}
\usepackage{amssymb}
\usepackage{color}
\usepackage{graphicx}
\usepackage{mathrsfs}
\usepackage{multirow}

\def\be{\begin{equation}}
\def\ee{\end{equation}}
\def\bea{\begin{eqnarray}}
\def\eea{\end{eqnarray}}
\def\no{\nonumber}

\newcommand{\omits}[1]{}

\begin{document}

\title{Entropy of Kerr-de Sitter black hole}
\author{Huai-Fan Li$^{a,b}$\footnote{Email: huaifan.li@stu.xjtu.edu.cn}, Meng-Sen Ma$^{a,b}$, Li-Chun Zhang$^{a,b}$ and Ren Zhao$^{b}$}

\medskip

\affiliation{\footnotesize$^a$Department of Physics, Shanxi Datong
University,  Datong 037009, China\\
\footnotesize$^b$Institute of Theoretical Physics, Shanxi Datong
University, Datong 037009, China}

\begin{abstract}
 Based on the consideration that the black hole horizon and the cosmological horizon of Kerr-de Sitter black hole are not independent each other, we conjecture the total entropy of the system should have an extra term contributed from the correlations between the two horizons, except for the sum of the two horizon entropies. By employing globally effective first law and effective thermodynamic quantities, we obtain the corrected total entropy and find that the region of stable state for kerr-de Sitter
is related to the angular velocity parameter $a$, i.e., the region of stable state becomes bigger as the rotating parameters $a$ is increases.

\end{abstract}

 \maketitle

\section{Introduction}
The astronomical observations show that our Universe is probably approaching
de Sitter spacetime. The cosmological constant corresponds to vacuum energy and is usually
considered as a candidate of dark energy. The accelerating universe
will evolve into another de Sitter phase. In order to construct the
entire history of evolution of the universe, we should have a clear
perspective to the classical and quantum properties of de Sitter
spacetime~\cite{Cai}. However, as is well known that in de Sitter space there is no
spatial infinity and no asymptotic Killing vector which is globally timelike.
Moreover, black holes in de Sitter spacetime cannot be in thermodynamic equilibrium
in general. Because there are multiple horizons with different temperatures for
de Sitter black holes.

The thermodynamic quantities
on the black hole horizon and the cosmological horizon all satisfy
the first law of thermodynamics, moreover the corresponding
entropies both fulfill the area formulae~\cite{Cai,Sekiwa,Miho,Taishi,Roy,
Ghezelbash,Gombero}. There have been accumulated interest to the thermodynamics
property of de Sitter spacetime in recent years~\cite{Dolan,Kamal,David,James,
Sourav,Pappas}.   In this way, the two horizons can be analyzed independently.
Besides, one can also take a global view to construct the globally effective
temperature and other effective thermodynamic quantities. No matter which
method is used, the total entropy of de Sitter black hole is supposed to
be the sum of both horizons, namely $S=S_++S_c$.~\cite{Myung,Zhao1,Zhao2}

We think that the correct method may to be so simple because the
event horizon and the cosmological horizon are not independent.
There may exist some correlations between them due to the following considerations.
We can take the Kerr-de Sitter black hole as example.
There are first laws of thermodynamics for both horizons.
According to \cite{Zhao5}, the thermodynamic quantities
on the two horizons satisfy the first law of thermodynamics system,
\bea\label{1st}
dM&=&T_{+}dS_{+}+\Omega_+ dJ+V_+d\Lambda,\no \\
dM&=&-T_{c}dS_{c}+\Omega_c dJ+V_cd\Lambda,
\eea
where $M$ is the conserved mass in de Sitter spacetime, $\Omega$ and $J$ stand for the angular velocities
and angular momentum, and the cosmological constant $\Lambda$ as a pressure and
define the thermodynamically conjugate variable to the thermodynamics volume, which can be
different from the geometrical volume, the subscript $''+''$ denotes the black hole horizon and the subscript $''c~''$ denotes the cosmological
horizon. Including the variable $\Lambda$, the above first laws can have corresponding Smarr formulae, which are also given in \cite{Cai2,Zhao6}.
The two laws in Eq.(\ref{1st}) are not truly independent. They depend on the same quantities $M,~J,~\Lambda$.
All the geometric and thermodynamic quantities for the both horizons can be represented by $M,~J,~\Lambda$. Therefore, the size of black hole horizon is closely related to the size of the cosmological horizon, and the evolution of black hole horizon will lead to the evolution of the cosmological horizon. Considering the
relation between the thermodynamic quantities on the two horizons is
very important for studying the thermodynamic properties of de
Sitter spacetime~\cite{Zhao3,Zhao4}.

Considering the correlation or entanglement between the black hole event horizon and the cosmological horizon,
the total entropy of the Kerr-de Sitter black hole is no longer simply $S=S_+ +S_c$, but should include an
extra term from the contribution of the correlations of the two horizons. In the Ref.~\cite{Ma}, we
calculate the extra term in the entropy for the spherically symmetric charged black hole in de Sitter spacetime. By the method we obtain the
reasonable interpretation for the R-N black hole in de Sitter spacetime. In this works, we will expand the method
to the axisymmetric rotating black hole in de Sitter spacetime. We can obtain the extra term form and
find the relates between the region of stable state and the rotating parameter $a$.

Our paper is organized as follows. In the next section we simple review the thermodynamics quantities of black hole horizon
and cosmological horizon of the Kerr-de Sitter black hole, obtain the condition that the effective temperature of the horizon
approaches to zero. In the base that the Kerr-de Sitter system satisfies the first laws of thermodynamics, the
entropy and the effect temperature of Kerr-de Sitter black hole is obtained, we study the condition that Kerr-de
Sitter system satisfies the stable equilibrium of the thermodynamics in Sec. {\ref{eff}}. Sec.{\ref{con}} is devoted
to conclusions.

\section{Kerr-de Sitter black hole}
Kerr-de Sitter black hole is a solution of Einstein equation in $(3 + 1)$
dimensions with a positive cosmological constant $\Lambda = \textstyle{3
\over {l^2}}$. It is characterised by parameters, mass $M$ and
angular momentum $J$. The spacetime metric of the Kerr-de Sitter black hole in
Boyer-Lindquist coordinates $(t,r,\theta,\varphi)$ is given by~\cite{Cai}
\be\label{staticmetric}
ds^2 = - \frac{\Delta _r }{\rho ^2}\left( {dt - \frac{a\sin ^2\theta }{\Xi
}d\phi } \right)^2 + \frac{\rho ^2}{\Delta _r }dr^2 + \frac{\rho ^2}{\Delta
_\theta }d\theta ^2 + \frac{\Delta _\theta \sin ^2\theta }{\rho ^2}\left(
{adt - \frac{r^2 + a^2}{\Xi }d\varphi } \right)^2,
\ee
where the various functions entering the metric are given by
$$
\rho ^2 = r^2 + a^2\cos ^2\theta ,
\quad
\Xi = 1 + \frac{\Lambda }{3}a^2,
\quad
\Delta _\theta = 1 + \frac{\Lambda }{3}a^2\cos ^2\theta ,
$$
\be\label{fr}
\Delta _r (r) = (r^2 + a^2)\left( {1 - \frac{\Lambda }{3}r^2} \right) -2mr,
\ee
the black hole horizon and cosmological horizon satisfies the equation $\Delta_r(r_+)=0$ and $\Delta _r (r_c ) =
0$, here $r_+$ and $r_c$ are the location of black hole horizon and cosmological horizon, respectively. So we can
obtain\cite{Zhao4}
\[
2m = \frac{(r_c + r_ + )(r_c^2 + a^2)(r_ + ^2 + a^2)}{r_c r_ + (r_c^2 + r_c
r_ + + r_ + ^2 + a^2)},
\]
\be
\label{eq3}
\Xi = \frac{r_c r_ + (r_c^2 + r_c r_ + + r_ + ^2 + 2a^2) - a^4}{r_c r_ +
(r_c^2 + r_c r_ + + r_ + ^2 + a^2)},\quad
\frac{\Lambda }{3} = \frac{r_c r_ + - a^2}{r_c r_ + (r_c^2 + r_c r_ + + r_ +
^2 + a^2)}.
\ee
In the de Sitter spacetime, by regarding respectively the black hole horizon and the cosmological horizon
as the thermodynamics systems, we can have~\cite{Sekiwa}
\[
T_ + = \frac{r_c^3 r_ + + r_c^2 r_ + ^2 - 2r_c r_ + ^3 + a^4 + 3a^2r_ + ^2 -
a^2(r_c^2 + r_ + ^2 + r_c r_ + + a^2)r_c / r_ + }{4\pi r_c (r_ + ^2 +
a^2)(r_c^2 + r_c r_ + + r_ + ^2 + a^2)},
\]
\be
\label{eq4}
T_c = - \frac{r_ + ^3 r_c + r_c^2 r_ + ^2 - 2r_ + r_c^3 + a^4 + 3a^2r_c^2 -
a^2(r_c^2 + r_c r_ + + r_ + ^2 + a^2)r_ + / r_c }{4\pi r_ + (r_c^2 +
a^2)(r_c^2 + r_c r_ + + r_ + ^2 + a^2)}.
\ee
The entropy and energy of the black horizon and the cosmological horizon are respectively,
\begin{equation}
\label{eq5}
S_ + = \frac{\pi (r_ + ^2 + a^2)}{\Xi }, \quad S_c = \frac{\pi (r_c^2 + a^2)}{\Xi },
\quad
M = \frac{m}{\Xi ^2}.
\end{equation}
The angular velocity of the black hole horizon and the cosmological horizon are respectively,
\begin{equation}
\label{eq6}
\tilde {\Omega }_ + = \frac{a\Xi }{r_ + ^2 + a^2},
\quad
\tilde {\Omega }_c = \frac{a\Xi }{r_c^2 + a^2},
\end{equation}
while the angular momentum and the angular velocity at infinity read
\begin{equation}
\label{eq7}
J = \frac{am}{\Xi ^2} = a\frac{(r_c + r_ + )(r_c^2 + a^2)(r_ + ^2 +
a^2)}{2r_c r_ + (r_c^2 + r_c r_ + + r_ + ^2 + a^2)\Xi ^2},
\quad
\Omega _\infty = \frac{\Lambda }{3}a.
\end{equation}
Taking the cosmological constant $\Lambda$ as the pressure, the thermodynamics quantities of
the both horizons satisfy the first laws of the thermodynamics~\cite{Sekiwa,Dolan},
\begin{equation}
\label{eq8}
\delta M = T_ + \delta S_ + + \Omega _ + \delta J + V_ + \delta P,
\end{equation}
\begin{equation}
\label{eq9}
\delta M = - T_c \delta S_c + \Omega _c \delta J + V_c \delta P,
\end{equation}
here
\begin{equation}
\label{eq10}
\Omega _ + = \tilde {\Omega }_ + - \Omega _\infty = \frac{a(1 - \Lambda r_ +
^2 / 3)}{r_ + ^2 + a^2}
 = \frac{a(r_c + r_ + )(r_c^2 + a^2)}{r_c (r_ + ^2 + a^2)(r_c^2 + r_ + ^2 +
r_c r_ + + a^2)},
\end{equation}
\begin{equation}
\label{eq11}
\Omega _c = \tilde {\Omega }_c - \Omega _\infty = \frac{a(1 - \Lambda r_c^2
/ 3)}{r_c^2 + a^2}
 = \frac{a(r_c + r_ + )(r_ + ^2 + a^2)}{r_ + (r_c^2 + a^2)(r_c^2 + r_ + ^2 +
r_c r_ + + a^2)}.
\end{equation}
The black hole and cosmological thermodynamics volumes are respectively~\cite{Dolan},
\begin{equation}
\label{eq12}
V_ + = \frac{r_ + A_ + }{3}\left[ {1 + \frac{a^2}{2\Xi r_ + ^2 }\left( {1 -
\frac{\Lambda }{3}r_ + ^2 } \right)} \right],
\end{equation}
\begin{equation}
\label{eq13}
V_c = \frac{r_c A_c }{3}\left[ {1 + \frac{a^2}{2\Xi r_c^2 }\left( {1 -
\frac{\Lambda }{3}r_c^2 } \right)} \right].
\end{equation}
From the Eq.(\ref{eq4}), when $a\ll r_+$, we can obtain the approximate value for the temperature
\[
T_ + = \frac{x + x^2 - 2x^3 + a^2(3x^2 - 1 / x - x - 1) / r_c^2 }{4\pi r_c
[(x^2 + a^2 / r_c^2 )(1 + x + x^2) + x^2a^2 / r_c^2 ]},
\]
\begin{equation}
\label{eq14}
T_c = - \frac{x^3 + x^2 - 2x + a^2(3 - x - x^2 - x^3) / r_c^2 }{4\pi r_c
x[(1 + a^2 / r_c^2 )(1 + x + x^2) + a^2 / r_c^2 ]}.
\end{equation}
where $x=r_+/r_c$.
\section{the effective thermodynamics of Kerr-de Sitter black hole}
\label{eff}
Generally, the temperatures at the black hole event horizon and cosmological horizon are not equal. Thus, the whole Kerr-de Sitter system cannot be in equilibrium thermodynamically.
However, there are two special cases for the Kerr-de Sitter black hole, in which the temperatures at the both horizons are the same. One case is the so-called Nariai black hole, the other is the lukewarm black hole~\cite{Dolan,Kamal,David,James,Sourav,Pappas}. For the Nariai black hole, the event horizon and cosmological horizon coincide \emph{apparently} and have the same temperature, zero or nonzero~\cite{Hawking}.

Considering the connection between the black hole horizon and the cosmological horizon~\cite{Zhao3,Zhao4}, we can derive the effective thermodynamic quantities and corresponding first law of black hole thermodynamics\footnote{One can also
take the cosmological constant $\Lambda$ as the pressure and then derive the effective volume. In this case the effective first law should be: $dM={T_{eff}}dS  + {V_{eff}}dP + {\Omega _{eff}}dJ$. This has been done in another  paper.}:
\be\label{eff1st}
dM = {T_{eff}}dS - {P_{eff}}dV + {\Omega _{eff}}dJ.
\ee
Here the thermodynamic volume is that between the black hole event horizon and the cosmological horizon, namely
\be\label{Vol}
V = V_c - V_ + \approx \frac{4\pi }{3}(r_c^3 - r_ + ^3 ) = \frac{{4\pi }}{3}r_c^3\left( {1 - {x^3}} \right).
\ee
The total entropy can be written as
\be\label{ent}
S = {S_ + } + {S_c} +S_{ex}= \pi r_c^2\left[1 + {x^2} + f(x)\right].
\ee
When $J$ is a constant, from $J=\frac{am}{\Xi^2}=aM $, we can obtain
\begin{equation}
\label{eq15}
\delta a = - \frac{a}{M}\delta M,
\end{equation}
\begin{equation}
\label{eq16}
\delta M = \left( {\frac{\partial M}{\partial r_c }} \right)_{x,a} \delta
r_c + \left( {\frac{\partial M}{\partial x}} \right)_{r_c ,a} \delta x +
\left( {\frac{\partial M}{\partial a}} \right)_{r_c ,x} \delta a,
\end{equation}
When $a\ll r_+$, from the eqs.(\ref{Vol}), (\ref{eq12}) and (\ref{eq13}), we can obtain
\begin{equation}
\label{eq18}
\delta V = \left( {\frac{\partial V}{\partial r_c }} \right)_x \delta r_c +
\left( {\frac{\partial V}{\partial x}} \right)_{r_c } \delta x
 = 4\pi r_c^2 (1 - x^3)\delta r_c - 4\pi r_c^3 x^2\delta x,
\end{equation}
From the Eq.(\ref{eff1st}), we can have the effective temperature of the Kerr-de Sitter black hole,
\bea
T_{eff} &=& \left( {\frac{\partial M}{\partial S}} \right)_{J,V} \no \\
 &=& \frac{A}{\left( {1 + \frac{a}{M}\left( {\frac{\partial M}{\partial
a}} \right)_{r_c ,x} } \right)\left[ {\left( {\frac{\partial S}{\partial x}}
\right)_{r_c ,a} \left( {\frac{\partial V}{\partial r_c }} \right)_x -
\left( {\frac{\partial S}{\partial r_c }} \right)_{x,a} \left(
{\frac{\partial V}{\partial x}} \right)_{r_c } } \right] - \frac{a}{M}\left(
{\frac{\partial S}{\partial a}} \right)_{r_c ,x} A}
\eea
here, $A=\left( {\frac{\partial M}{\partial x}} \right)_{r_c ,a} \left(
{\frac{\partial V}{\partial r_c }} \right)_x - \left( {\frac{\partial
M}{\partial r_c }} \right)_{x,a} \left( {\frac{\partial V}{\partial x}}
\right)_{r_c } $. When $a\ll r_+$, $\Xi\sim 1$. we can obtain via neglecting the higher order of $a$,
\begin{equation}
\label{eq19}
T_{eff} = \left( {\frac{\partial M}{\partial S}} \right)_{J,V}
 \approx \frac{\left( {\frac{\partial M}{\partial x}} \right)_{r_c ,a}
\left( {\frac{\partial V}{\partial r_c }} \right)_x - \left( {\frac{\partial
M}{\partial r_c }} \right)_{x,a} \left( {\frac{\partial V}{\partial x}}
\right)_{r_c } }{\left[ {\left( {\frac{\partial S}{\partial x}} \right)_{r_c
,a} \left( {\frac{\partial V}{\partial r_c }} \right)_x - \left(
{\frac{\partial S}{\partial r_c }} \right)_{x,a} \left( {\frac{\partial
V}{\partial x}} \right)_{r_c } } \right]},
\end{equation}
From the eq. (\ref{eq3}), we have
\begin{equation}
\label{eq20}
M \approx \frac{(r_c + r_ + )(r_c^2 + a^2)(r_ + ^2 + a^2)}{2r_c r_ + (r_c^2
+ r_c r_ + + r_ + ^2 + a^2)} \approx r_c \frac{(1 + x)(x^2 + (1 + x^2)a^2 /
r_c^2 )}{2x(1 + x + x^2)}.
\end{equation}
Considering the expression for the entropy of both horizons (\ref{ent}), we can
take the entropy of Kerr-de Sitter black hole as
\begin{equation}
\label{eq21}
S = \frac{\pi }{\Xi }r_c^2 \left( {1 + x^2 + f(x) + 2\frac{a^2}{r_c^2 }}
\right) \approx \pi r_c^2 \left( {1 + x^2 + f(x) + 2\frac{a^2}{r_c^2 }}
\right),
\end{equation}
Here the undefined function $f(x)$ represents the extra contribution from the correlations of the two horizons.
Then, how to determine the function $f(x)$?
Inserting eqs. (\ref{eq21}), (\ref{eq20})and (\ref{Vol}) into eq.(\ref{eq19}), we can obtain
\begin{equation}
\label{eq25}
T_{eff} = \frac{1}{2\pi r_c }\frac{T_1 (x)}{T_2 (x)},
\end{equation}
where
$$
T_1 (x) = \frac{1}{x^2(1 + x + x^2)}\left[ {x^2(1 + 2x)(1 - x) + x^5(1 + x)
- \frac{a^2}{r_c^2 }\left( {(1 + x^4)(1 + x + x^2) - 2x^3} \right)}
\right],
$$
\begin{equation}
\label{eq26}
T_2 (x) = \left[ {2x(1 + x) + 2x^2f(x) + (1 - x^3)f'(x)} \right].
\end{equation}
Neglecting the higher order terms of $a$, we can rewrite the eq.(\ref{eq4}),
$$
T_ + = \frac{x + x^2 - 2x^3 + a^2(3x^2 - 1 / x - x - 1) / r_c^2 }{4\pi r_c
[(x^2 + a^2 / r_c^2 )(1 + x + x^2) + x^2a^2 / r_c^2 ]},
$$
\begin{equation}
\label{eq27}
T_c = - \frac{x^3 + x^2 - 2x + a^2(3 - x - x^2 - x^3) / r_c^2 }{4\pi r_c
x[(1 + a^2 / r_c^2 )(1 + x + x^2) + a^2 / r_c^2 ]}.
\end{equation}
From the solution of (\ref{eq27}), when
\be
\label{arc}
\frac{a^2}{r_c^2 } = \frac{x^4 + 5x^3 + 5x^2 + 5x + 1 - \sqrt {x^8 + 10x^7 +
31x^6 + 56x^5 + 69x^4 + 56x^3 + 31x^2 + 10x + 1} }{2\left( {x^2 + 1}
\right)},
\ee
the radiation temperature of the both horizons is equal.
\begin{figure}[!htbp]
\center{
\includegraphics[width=7cm,keepaspectratio]{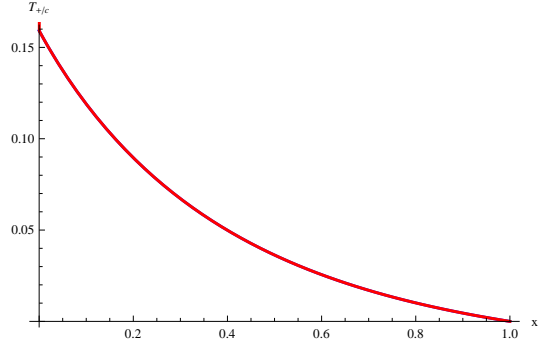}
\caption{$T_{+/c}$ with respect to $x$. When $a^2/r_c^2$ satisfies the condition of (\ref{arc}),
$T_+$ and $T_c$ is equal. }}
\end{figure}

In general cases, the temperatures of the black hole horizon and the cosmological horizon are not the same, thus the globally effective temperature $T_{eff}$ cannot be compared with them. However, in the special case, such as lukewarm case~\cite{Mellor,Romans}, the temperatures of the two horizons are the same. We conjecture that in this special case the effective temperature should also take the same value. On the basis of this consideration, we can obtain the information of $f(x)$.
For Kerr-de Sitter black hole,
\begin{equation}
\label{eq28}
T_{eff} = \tilde {T}_c = \tilde {T}_ + .
\end{equation}
From the eq.(\ref{eq25}) and (\ref{eq28}), we can obtain
\begin{equation}
\label{eq29}
T_{eff} = \frac{T_1 (x)}{\tilde {T}_1 (x)}\tilde {T}_ + [\tilde {T}_c ],
\end{equation}
where $\tilde {T}_1 (x)$ is the value of $T_1 (x)$ with $a^2/r^2_c$ in eq.(\ref{arc}).
\begin{figure}[!htbp]
\center{
\includegraphics[width=7cm,keepaspectratio]{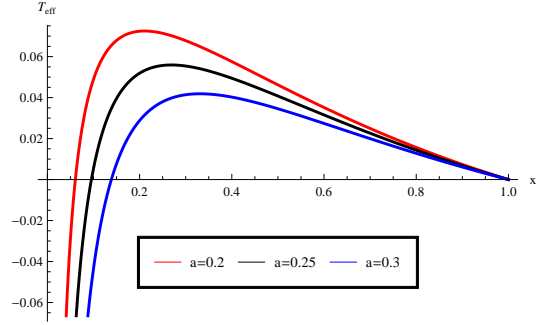}
\caption{$T_{eff}$ with respect to $x$. $T_{eff}$ has a maximum at $x=0.21$ for $a=0.2$,
$x=0.269$ for $a=0.25$, $x=0.331$ for $a=0.3$.\label{tex}}
}
\end{figure}
Substuting the eqs.(\ref{eq25}) and (\ref{eq28}) into the eq (\ref{eq29}), we can obtain
\begin{equation}
\label{eq30}
T_2 (x) = \frac{1}{2\pi r_c }\frac{\tilde {T}_1 (x)}{\tilde {T}_c [\tilde
{T}_ + ]},
\end{equation}
Inserting eqs.(\ref{eq26}) and (\ref{eq27}), we can obtain the equation of $f(x)$
\bea \label{eq31}
2x(1 + x) + 2x^2f(x) + (1 - x^3)f'(x)
 + 2[(1 + a^2 / r_c^2 )(1 + x + x^2) + a^2 / r_c^2 ] \no \\ \frac{x^2(1 - x)(1 + 2x)
+ x^5(1 + x) - \left[ {(1 + x^4)(1 + x + x^2) - 2x^3} \right]a^2 / r_c^2
}{x(1 + x + x^2)\left[ {x^3 + x^2 - 2x + a^2(3 - x - x^2 - x^3) / r_c^2 }
\right]}=0
\eea
\begin{figure}[!htbp]
\center{
\includegraphics[width=7cm,keepaspectratio]{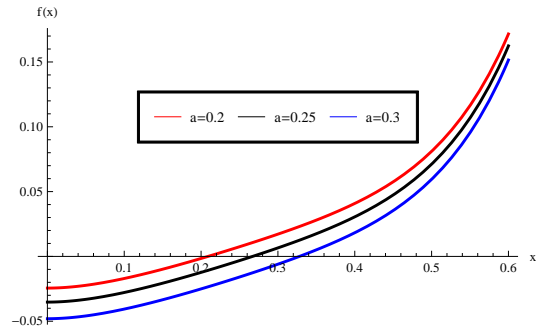}
\caption{$f(x)$ with respect to $x$. $f(x)$ is zero at $x=0.21$ for $a=0.2$,
$x=0.269$ for $a=0.25$, $x=0.331$ for $a=0.3$. \label{fx}}
}
\end{figure}

\begin{figure}[!htbp]
\center{
\includegraphics[width=7cm,keepaspectratio]{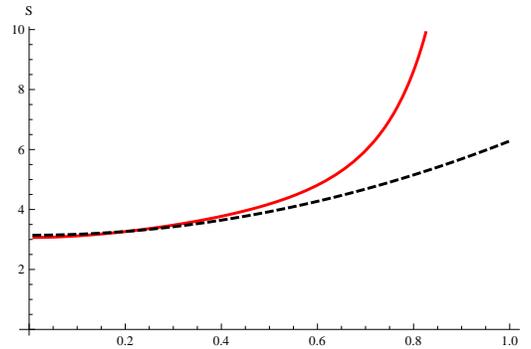}
\caption{$S$ with respect to $x$. the dashed(black) curve represents the sum of the
two horizon entropy and the solid(red) curve depicts the result in Eq.(\ref{eq21}). In the calculation we set $r_c=1$.\label{sx}}
}
\end{figure}

In Fig.\ref{tex}, \ref{fx} and \ref{sx}, we depict the effective temperature $T_{eff}$, $f(x)$ and $S$ as functions of $x$. It is shown that $T_{eff}$ tends to zero as $x \rightarrow 1$, namely the Nariai limit. Although this result does not agree with that of Bousso and Hawking~\cite{Hawking}, it is consistent with the entropy. As is depicted in Fig.\ref{sx}, the entropy will diverge as $x \rightarrow 1$.

The specific heat of kerr-de Sitter system can be defined as
\begin{equation}
\label{eq32}
C_{r_c ,a} = T_{eff} \left( {\frac{\partial S}{\partial T_{eff} }}
\right)_{r_c ,a} .
\end{equation}

\begin{figure}[!htbp]
\center{
\includegraphics[width=7cm,keepaspectratio]{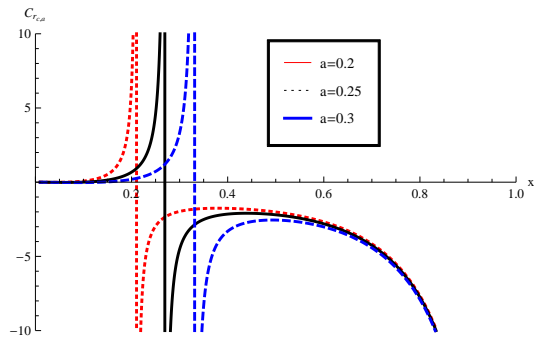}
\caption{$C_{r_c,a}$ with respect to $x$. In the calculation we set $r_c=1$.\label{cx}}
}
\end{figure}

From Fig.\ref{cx}, when $0<x<x_c$, the specific heat of system is positive, while $x_c<x<1$, it is negative. This
means that the Kerr-de Sitter black hole with $x<x_c$ is thermodynamically stable. From the Fig.\ref{cx},
we can find that the region of stable state for kerr-de Sitter
is related to the angular velocity parameter $a$, i.e., the region of stable state becomes bigger as the rotating parameters $a$ is increases.

\section{conclusion}
\label{con}
In this letter, we mainly studied the entropy of Kerr-de Sitter black hole. In addition, we analyzed the thermodynamic stability
of this black hole. Firstly, we simply review the thermodynamic quantities of black hole horizon and cosmological
horizon of the Kerr-de Sitter black hole. Considering the relation of the black hole horizon and
the cosmological horizon, we conjecture the total entropy should take the form of Eq.(\ref{ent}). Secondly, we find that
the entropy obtained by Eq.(\ref{ent}) monotonically increases as $x$ increased. When $x\rightarrow1$, this entropy diverges.
Finally, it is found that the Kerr-de sitter black hole is unstable due to negative heat capacity in some region for rotating parameter $a$.
In a word, we obtain the corrected entropy of Kerr-de Sitter black hole, which may make the method more acceptable.

\section*{Acknowledgements}
We would like to thank Dr Hui-Hua Zhao and Yu-Bo Ma for their indispensable discussions and comments. This work was supported by the Young Scientists Fund of the National Natural Science Foundation of China (Grant No.11205097), in part by the National Natural Science Foundation of China (Grant No.11475108), Supported by Program for the Innovative Talents of Higher Learning Institutions of Shanxi, the Natural Science Foundation of Shanxi Province,China(Grant No.201601D102004) and the Natural Science Foundation for Young Scientists of Shanxi Province,China (Grant No.2012021003-4), the Natural Science Foundation of Datong city(Grant No.20150110).

\end{document}